# Sensor fusion luminescence thermometry – better together

Aleksandar Ćirić[*], Zoran Ristić, Tamara Gavrilović, Jovana Periša, Mina Medić, Miroslav D. Dramićanin

*Centre of Excellence for Photoconversion, Vinča Institute of Nuclear Sciences – National Institute of the Republic of Serbia, University of Belgrade, P.O. Box 522, Belgrade 11001, Serbia*

*Corresponding author: aleksandar.ciric@ff.bg.ac.rs

**Abstract**

Luminescent thermometers are highly effective in niche applications such as nano-thermometry, *in vivo* imaging, and extreme conditions like high electromagnetic fields, radiation, and under mechanical or chemical stress. Advancing measurement precision, temperature resolution, and extending the temperature range are the main problems in this rapidly evolving research field, crucial in the work of R&D industry engineers and scientific researchers. Traditionally using single parameter for sensing leads to the sensor underperformance. Sensor fusion (SF), a well-established statistical tool used to enhance precision in autonomous vehicles, navigation systems, medical imaging, and wearable devices is here engineered for luminescent thermometry, by either fusing multiple sensor probes, or observing each temperature dependent feature as a separate sensor, resulting in always increased precision and extended usable temperature range. SF is a replacement and superior alternative to multiple linear regression models as it stands out for its versatility, enabling performance limits to be reached with any sensor probe material. It is compatible with both time-resolved and steady-state readouts, used independently or in combination, with single or multiple excitations. The benefits are illustrated using probes with $Mn^{5+}$, $Yb^{3+}/Er^{3+}/Ho^{3+}$, $Sm^{2+}$, and $Mn^{4+}/Ho^{3+}/Cr^{3+}$ activator ions.

1. Introduction

   1.1. Luminescence thermometry: perspectives and applications

In 2024, the temperature sensors market, valued at 8.8 billion USD, is driven by extensive applications across various industries, including consumer electronics, automotive, healthcare, and industrial sectors, with a projected growth to 12 billion USD by the end of the decade [1]. Most sensors are contact sensors like thermistors, thermocouples, or RTDs. The market has seen a significant rise in the popularity of IR sensors, especially during the COVID-19 pandemic, though they were already well-established in the industrial sector where contactless measurements or surface mapping are needed. While IR and contact sensors suffice for most applications, advancements in science and technology have led to the development of the alternative optical temperature sensing methods - luminescence thermometry, unbeatable at various niche applications [2].

Luminescence thermometry leverages the temperature-dependent properties of luminescent materials, where temperature readout is achieved by monitoring emission features like changes in luminescence lifetime, phase-shift, spectral shape, or shifts of emission lines [3]. This method can achieve high precision and sensitivity, often surpassing traditional sensing technologies, especially when the luminescent properties of the material are extremely sensitive to temperature changes [4]. It provides high spatial resolution temperature mapping, particularly useful in micro-scale and nano-scale applications where detailed temperature profiles are needed. Similar to IR thermometry, it allows for remote sensing, which is crucial for moving or microscopic objects. Unlike IR thermometry, it can be applied from cryogenic temperatures up to 1700°C [5], making it versatile for various applications. While IR thermometers are non-contact, they are limited by surface emissivity and may not provide accurate readings in reflective or semi-transparent materials. Luminescence thermometry, however, can measure through transparent media and provide more reliable data [6]. Contact thermometers are limited by response time, accuracy in extreme conditions, and invasiveness. Luminescence thermometry is only semi-invasive, making it suitable for measuring temperature within living microorganisms and unaffected by environmental conditions, allowing accurate measurements in high electromagnetic or radioactive fields or corrosive environments [7,8].

The niche applications of luminescence thermometry include microelectronics, materials science, medical diagnostics, *in vivo* thermal imaging, gas turbines and jet engines, spacecraft, understanding biological processes of microorganisms, cryogenics, nanotechnology, and monitoring chemical processes [9]. In microelectronics, it is used during fabrication and testing where precise temperature mapping at micro- and nano-scale is critical [10]. In materials science, it is used for studying the thermal properties of new materials. Luminescent nanoparticles can be used for temperature mapping inside biological tissues, crucial for hyperthermia treatment of cancer, where precise temperature control is required to ensure effective treatment without damaging surrounding healthy tissues [11–13]. *In vivo* imaging of temperature distributions within organisms can help understand various physiological processes and treatment effects [14–16].

Luminescence thermometry is particularly useful in extreme conditions where traditional sensors might fail, such as in the combustion chambers of gas turbines and jet engines or in spacecraft where high radiation and wide temperature ranges are encountered [17]. It is also employed in cryogenic research, where traditional thermometers may not provide accurate readings at extremely low temperatures. As technology scales down, traditional temperature measurement methods become less effective, and luminescence thermometry can provide the necessary precision at the nano-scale [18]. In temperature-sensitive chemical reactions requiring non-intrusive measurements, luminescence thermometry offers a solution without interfering with the process [19].

Despite its numerous advantages and versatile applications, luminescence thermometry has yet to see widespread usage. Improvements in precision, speed, and temperature range are necessary for broader adoption. Current research trends focus on enhancing sensor probe materials and readout methods [20]. The investigation into materials to find luminescent emissions highly sensitive to temperature change has matured, with significant advancements in luminescent nanoparticles, quantum dots, single crystals, films, and coatings. However, it was recently demonstrated that selecting single temperature dependent parameter does not exploit the full potential of the sensor probe material [21]. Improvements in readout methods have been slower, with recent advances in using multiple parameters such as dimensionality reduction methods and multivariable analysis [22,23]. Future research will likely focus on combining superior materials with advanced readout techniques to unlock the full potential of luminescence thermometry [24], however the multivariable analysis has several issues that limit its wider application.

### 1.2. Sensor accuracy and precision

Temperature sensors do not directly provide the temperature value itself; instead, they measure another quantity (parameter - indication, Δ) that is then converted to the temperature scale. These sensors consist of two main components: the sensor probe material and the parameter readout device. The quality of any sensor, defined by its accuracy and precision, depends heavily on both components (see Figure 1).

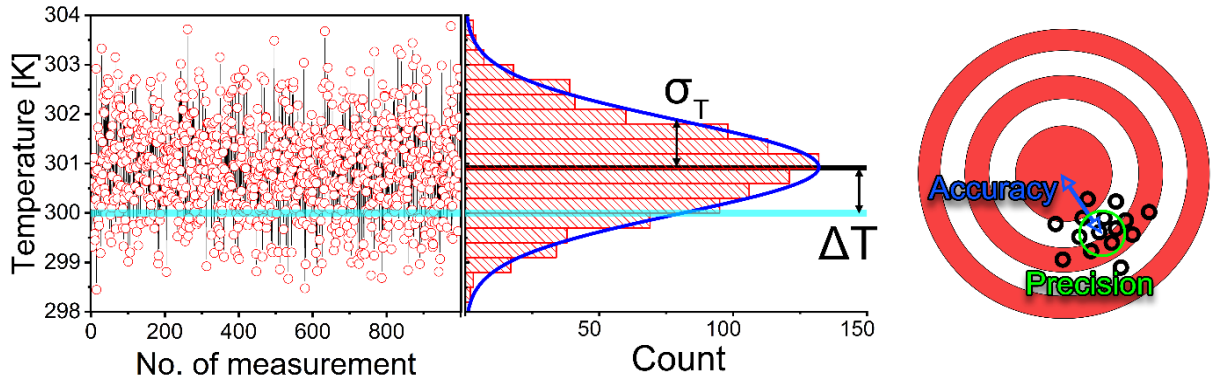

Figure 1. Simulated 1000 measurements of temperature with example real value of 300 K (cyan line) and the corresponding histogram and Gaussian probability distribution of the measured values. Accuracy describes how close the mean of the measured values is from the real value, while precision is the measure of the spread, given by standard deviation. An illustrative example of those two terms with shooting at the target is given in the image to the right.

The calibration curve that links the parameter to temperature is expressed as a function *T(Δ)*. This function *T(Δ)* can be viewed as a Gaussian distribution of measurements, with its mean and standard deviation dependent on the temperature: $\mathcal{N}(T, \sigma)$.

The temperature estimated from *T(Δ)* may differ from the actual temperature of the object. Accuracy is a measure of how close a measurement is to the true or accepted value, while precision indicates the consistency or repeatability of the measurements. In essence, accuracy reflects correctness, and precision reflects the closeness of repeated measurements to each other. It is possible for measurements to be precise but not accurate, accurate but not precise, neither, or both.

Accuracy is related to systematic error, which can often be corrected through calibration. For instance, adjusting the iron sights of a gun when shooting at a target improves accuracy. Precision, however, relates to statistical error, indicating the variability or spread of repeated measurements. Usually, one type of error tends to be more dominant. Systematic error can be significant if there is a consistent bias in the measurement system. Therefore, good measuring systems are designed to be highly accurate [25].

### 1.3. Sensor fusion

Sensor fusion (SF) combines data from multiple sensors to provide more accurate, reliable, and comprehensive information than any single sensor alone. This technique is essential in fields such as robotics, autonomous vehicles, aerospace, medical diagnostics, and smart environments. Its main goal is to enhance information quality by reducing uncertainty and improving system robustness. Involving multiple sensors measuring the same parameter to cross-verify data reduces errors and increases reliability [26].

In autonomous vehicles, sensor fusion merges data from cameras, radar, LIDAR, and ultrasonic sensors to perceive surroundings and make driving decisions [27]. In robotics, it integrates data from vision systems, tactile sensors, and proprioceptive sensors for effective navigation and object manipulation. In aerospace, it combines data from gyroscopes, accelerometers, and GPS to enhance navigation and control accuracy in aircraft and spacecraft. In medical diagnostics, it fuses data from various imaging modalities and physiological sensors for comprehensive diagnosis. In smart homes and IoT, it uses data from sensors like temperature, humidity, and motion detectors to create intelligent, responsive environments [28]. It is even used in our mobile phones to obtain the precise location by combining information from multiple GPS satellites, Wi-Fi or mobile networks [29].

SF integrates measured data from various sources into a unified piece of information. If Gaussian distributions $\mathcal{N}(T_i, \sigma_i)$ represent each parameter, $\Delta_i$, SF combines them to produce a fused estimate, which is also expressed as a Gaussian distribution (see Figure 2) [30]. This process consistently results in reduced variance, leading to increased precision and enhanced resolution. However, it does not necessarily improve accuracy. This is not a limitation of SF but underscores the importance of proper calibration for all sensors involved in the process.

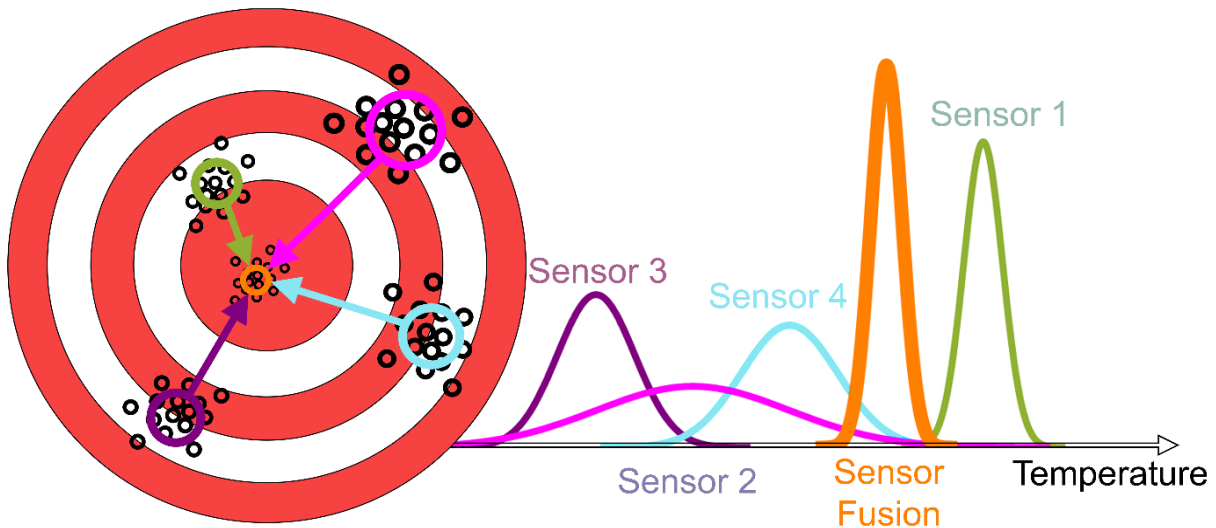

Figure 2. Sensor Fusion accuracy and measurement calculated from 4 independent sensors.

This research investigates the application of SF in luminescence thermometry, providing clear guidelines to enhance precision, expand the sensor temperature range, or achieve both. It explains how multiple temperature-dependent features in steady-state spectra can be simultaneously utilized, as well as multiple readouts in lifetime or frequency domain methods. These concepts are tested on various probe materials, demonstrating through examples that SF theory is effective in practical luminescence thermometry and that it solves the issue with underperformance by using a single parameter for temperature sensing. A detailed comparison with another multiparametric method, multiple linear regression model (MLR) is made, demonstrating that SF successfully overcomes the problems of the MLR supervised machine learning algorithm.

## 2. Experimental

Synthesis and phase characterizations of all investigated samples are given in the supplementary information.

Photoluminescence (PL) emission spectra of $Mn^{5+}$ doped sample were recorded over a temperature range from 20 °C to 100 °C using a custom Peltier-based heating stage described in Ref. [31]. The excitation source was an Ocean Insight 635 nm fiber-coupled LED. The PL emission spectra were captured using an Ocean Insight NIRQuest+ spectrometer connected via a bifurcated fiber-optic cable. Temperature of other samples was controlled by the MikroOptik stage. Steady-state spectra were recorded via Ocean Optics FX spectrometer, except for the $Yb^{3+}/Er^{3+}$ sample where a FHR 1000 high-resolution spectrograph with an Horiba Jobin-Yvon ICCD detector was used. Lifetimes were recorded via Hamamatsu H10722-20 PMT. Excitation and emission signals were passed through a bifurcated fiber-optic cable. $Sm^{2+}$, $Cr^{3+}$, $Mn^{4+}$, and $Ho^{3+}$ doped samples were excited by fiber-coupled 450 nm laser. $Yb^{3+}/Er^{3+}$ sample was irradiated by 980 nm laser. The separate emissions for lifetime measurements were selected by the bandpass filters with 10 nm full-width-half-maximum.

## 3. Results & Discussion
### 3.1. Sensor Fusion Luminescence Thermometry: Theory
#### 3.1.1. Luminescence thermometry: parameters, uncertainty, resolution, precision

The luminescence of a substance varies with temperature, manifesting through multiple phenomena that form the foundation for temperature measurement methods. These phenomena underpin the principles of luminescence thermometry, which can be categorized into two main techniques based on the temporal characteristics of the luminescent changes: time-integrated (steady-state) and time-resolved. Time-integrated methods involve constant illumination and continuous observation, while time-resolved methods use pulsed or harmonically modulated excitation. Methods for time-integrated measurements include luminescence intensity ratio (LIR), bandwidth (Γ), and peak energy shifts (ν), while time-resolved measurements utilize luminescence lifetime (τ) and phase-shift (φ) [32]. Traditionally, only the parameter with the highest sensitivity is chosen for sensing, ignoring other parameters.

In conventional thermometry an parameter Δ(T) is measured, and the uncertainty of measurement Δ is given as the standard deviation $\sigma_\Delta$. The absolute sensitivity of the parameter is given by the rate of change of the parameter with temperature [33]:

$$S_\Delta = \left|\frac{\partial \Delta}{\partial T}\right| \qquad (1)$$

then the uncertainty in measured temperature is calculated from the sensitivity and uncertainty in parameter [34]:

$$\sigma = \left|\frac{\partial T}{\partial \Delta}\right|\sigma_\Delta = \frac{\sigma_\Delta}{S_\Delta} \qquad (2)$$

The distinction between uncertainty in measured temperature and temperature resolution is crucial yet often misunderstood. Temperature resolution refers to the smallest change in temperature that an instrument can reliably detect and display. In contrast, the standard deviation quantifies the repeatability and reliability of the measurements. While the standard deviation measures variability, temperature resolution assesses the instrument's sensitivity and its ability to detect minute changes in temperature. To ensure that measurements are both reliable and distinguishable from noise, it is common practice to set the resolution threshold at a level where the signal can be confidently separated from background noise. Although some sources equate resolution to the standard deviation [35], a higher confidence level is generally preferable. For instance, setting the resolution to twice the standard deviation (2σ) corresponds to a 95% confidence level. Hence, the temperature resolution is more accurately defined as [36]:

$$\Delta T[K] = 2\sigma \qquad (3)$$

Finally, the precision of a sensor is defined as the inverse of its variance, and this is the most intuitive measure for comparison of different sensors [37]:

$$p[K^{-2}] = \frac{1}{\sigma^2} \qquad (4)$$

### 3.1.2. Sensor Fusion Luminescence Thermometry

There are three primary approaches to performing SF luminescence thermometry: (i) combining measurements from multiple sensors, (ii) treating each measurement as an independent sensor, or (iii) a combination of both approaches. SF luminescence is versatile and not confined to time-resolved or time-integrated modes; it can be simultaneously applied to various parameters such as line-shift, lifetime, or phase-shift, as shown in Figure 3.

When fusing different parameters or sensor responses, the initial step is to reverse-map the data. While measurements provide $\Delta_i(T)$, SF thermometry requires determining the inverse function $T_i = T(\Delta_i)$. Each parameter $\Delta_i$ has a unique associated uncertainty, meaning that $T_i$ will also have its own uncertainty, $\sigma_i$. To estimate the combined result, each measurement's contribution is weighted according to its uncertainty. The weights are inversely proportional to the variance, indicating that individual weights are directly related to their precision:

$$p_i = \frac{1}{\sigma_i^2} \qquad (5)$$

The fused variance in SF is then equal to the inverse of the sum of weights [38]:

$$\sigma^2 = \frac{1}{\sum_i p_i} \qquad (6)$$

σ is the uncertainty of the SF calculation, given as the standard deviation in fused Gaussian distribution, $\mathcal{N}(T, \sigma)$.

The fused estimate of temperature is calculated from the temperature estimates of each sensor, weighed by their precisions [38]:

$$T = \frac{\sum_i p_i T_i}{\sum_i p_i} = \sigma^2 \sum_i \frac{T_i}{\sigma_i^2} \qquad (7)$$

Ultimately, the precision of the fused temperature estimate is equal to the sum of precisions of individual parameters, proving that SF always leads to better sensing:

$$p = \frac{1}{\sigma^2} = \sum_i p_i \qquad (8)$$

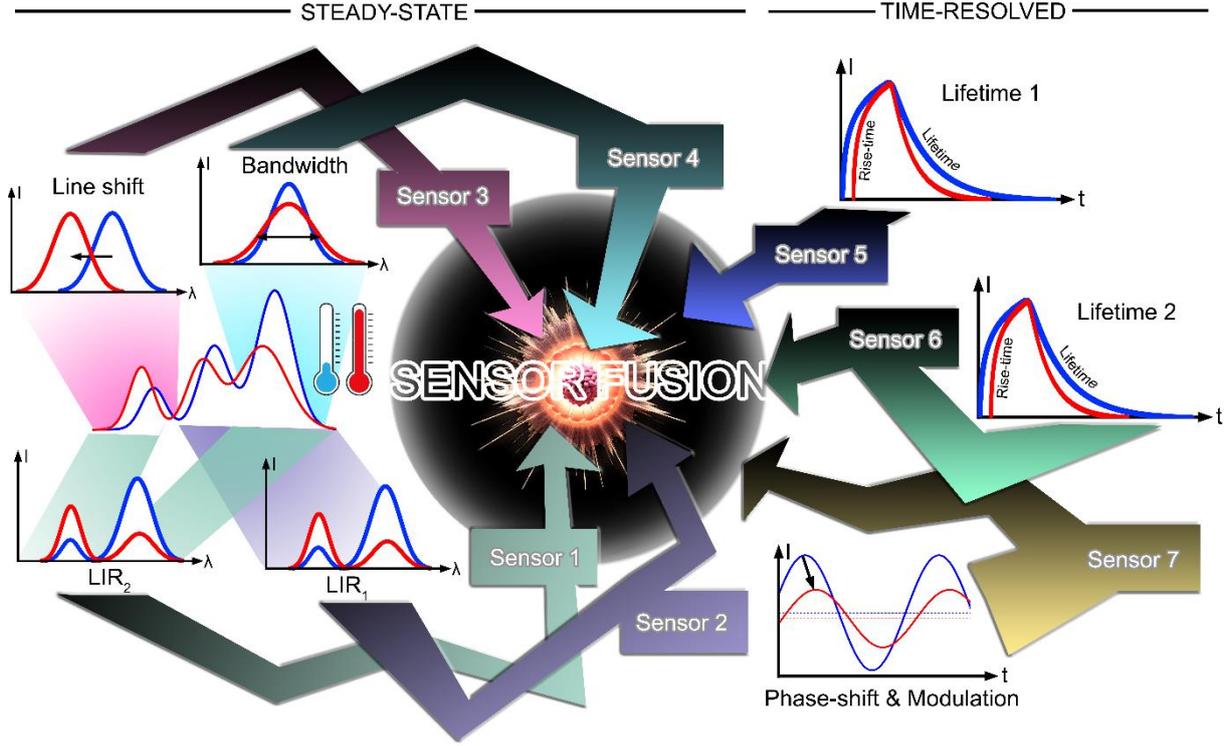

Figure 3. Sensor fusion in luminescence thermometry on an example of 7 sensors – parameters extracted from the temperature-dependent spectra or time-resolved measurements: (1) $LIR_1$, (2) $LIR_2$, (3) Line shift, (4) Bandwidth, (5) Lifetime 1, (6) Lifetime 2, (7) Phase-shift and Modulation.

Note: for fused estimate of uncertainty uncertainties in temperature of each parameter, $\sigma_i$, must be used instead of uncertainties in parameter, $\sigma_{\Delta i}$. Uncertainties in temperature can be directly obtained by calculating the standard deviation of $T_i$ or indirectly from the uncertainties in parameters (which are obtained by the standard deviation of $\Delta_i(T)$):

$$\sigma_i(T) = \left|\frac{\partial T_i}{\partial \Delta_i}\right| \sigma_{\Delta_i} \tag{9}$$

which is easily calculated from $\Delta_i(T)$. The inverse sensitivity that links uncertainty of parameter with uncertainty in temperature for the example of the Mott-Seitz (Arrhenius) relation for temperature dependent lifetime is given by:

$$\tau(T) = \frac{\tau_0}{1 + A \exp\left(-\frac{\Delta E}{kT}\right)} \Rightarrow \frac{\partial T(\tau)}{\partial \tau} = \frac{\Delta E \tau_0}{k\,\tau(\tau - \tau_0) \ln^2 \frac{A\tau}{\tau_0 - \tau}} \tag{10}$$

### 3.1.3. Comparison with MLR

Multiple Linear Regression (MLR) is a statistical method used to model the relationship between a dependent variable and multiple independent variables, assuming a linear relationship between them. The model is expressed as:

$$T = \beta_0 + \beta_1 \Theta_1 + \cdots + \beta_n \Theta_n + \varepsilon \tag{11}$$

where $T$ is the dependent variable, $\Theta_i$ are the independent variables, $\beta_i$ are the coefficients determined by fitting the model, and $\varepsilon$ is the error term. MLR extends the least squares method to multiple dimensions, minimizing the residuals between the predicted and actual values of the dependent variable by minimizing the sum of the squared residuals.

MLR requires a linear relationship between variables [21], which may not always hold in real-world data. One way to address this is by mapping the temperature readouts to different independent variables, $\Theta_i = T(\Delta_i) = T_i$, which can account for non-linearity. However, this introduces dependency among the variables — since all sensors measure the same temperature, the $T_i$ values are interdependent. As a result, MLR tends to prioritize the most accurate sensor readings rather than a balance among them. This differs from SF, which focuses on reliability and precision, as SF by design favors more reliable measurements and does not impose the independency requirements.

MLR is also sensitive to outliers, which can significantly affect the model. In contrast, SF with invariance weighting assigns weights to sensor readings based on their reliability or invariance, producing a more robust combined measurement. Unlike MLR, SF does not assume a specific functional form and focuses on optimally combining sensor data based on individual sensor characteristics.

### 3.2. Sensor Fusion Luminescence Thermometry: Practice

Steps for obtaining SF:

1. Perform conventional thermometry to obtain dependence of parameters $\Delta_i(T)$, where $i$ represents parameters like LIR, line-shift, bandwidth, lifetime, phase-shift etc. Measure each $\Delta_i$ multiple times at the same temperature to estimate its standard deviation. Repeat those measurements at each measured temperature, as standard deviation is not temperature invariant. Ensure acquisition times for each parameter are the same to properly evaluate uncertainties.
2. Fit the experimental data $\Delta_i(T)$ to the appropriate function.
3. Find the inverse of the previously fitted function $\Delta_i(T) \to T(\Delta_i) = T_i$. Common examples include:

    - Boltzmann LIR: $R(T) = B e^{-\frac{\Delta E}{kT}} \Rightarrow T(R) = \frac{\Delta E}{k \ln\frac{B}{R}}$
    - Arrhenius lifetime: $\tau(T) = \frac{\tau_0}{1 + A \exp\left(-\frac{\Delta E}{kT}\right)} \Rightarrow T(\tau) = \frac{\Delta E}{k \ln\frac{A\tau}{\tau_0 - \tau}}$
    - Varshni line-shift [39]: $\Delta E = E(0\,K) - E(T) = \frac{\alpha T^2}{\beta + T} \Rightarrow T = \frac{\Delta E + \sqrt{\Delta E^2 + 4\alpha\beta\Delta E}}{2\alpha}$

    For a complete list of temperature dependent luminescent models see Ref. [32].
4. Estimate $\sigma_i$ are estimated as the standard deviation of $T(\Delta_i)$.
5. Calculate the fused uncertainty from individual uncertainties using: $\sigma^2 = \left(\sum_i \frac{1}{\sigma_i^2}\right)^{-1}$
6. Perform the fused temperature estimation by: $T = \sigma^2 \sum_i \frac{T_i}{\sigma_i^2}$
7. (Optional) To validate the SF method, ensure the standard deviations of $T$ in the previous step are equal to the standard deviations in step 5 for each measured temperature.

Potential experimental or practical setups for various readouts are illustrated in Figure 4. Although fiber-optic setups are shown for demonstrative purposes, SF luminescence thermometry can be implemented with any optical apparatus. Phase-shift or Lifetime measurements can be detected using a pulsed excitation source and a detector. LIR methods require monitoring at least two wavelengths, which means two detectors with filters are needed. Multilevel-Cascade LIR (McLIR), which uses multiple LIR readouts, requires as many detectors as there are emissions observed [40]. Single-band ratiometric (SBR) readout observes a single emission while switching excitations [41]. $LIR^2$ is a combination of LIR and SBR, requiring two excitation sources and two detectors [42]. Both $LIR^2$ and McLIR are designed to achieve greater precision at elevated temperatures; however, they require algorithms that switch the observations of parameters depending on the temperature to the parameter with the highest precision. All single-parametric methods except SBR and $LIR^2$ can be employed with a spectrometer and a single excitation source.

In contrast, SF does not require switching algorithms. In a setup with a pulsed excitation source and multiple detectors, SF can combine all time-resolved measurements with LIR. When combined with a spectrometer, all steady-state measurements can be fused. Using only a spectrometer, all steady-state parameters can be fused. If the spectrometer works with time dependence, SF requires just one detection system to employ all possible parameters. Ultimately, SF does not necessarily need to employ a single probe; it can combine multiple probes with various excitations and detection systems.

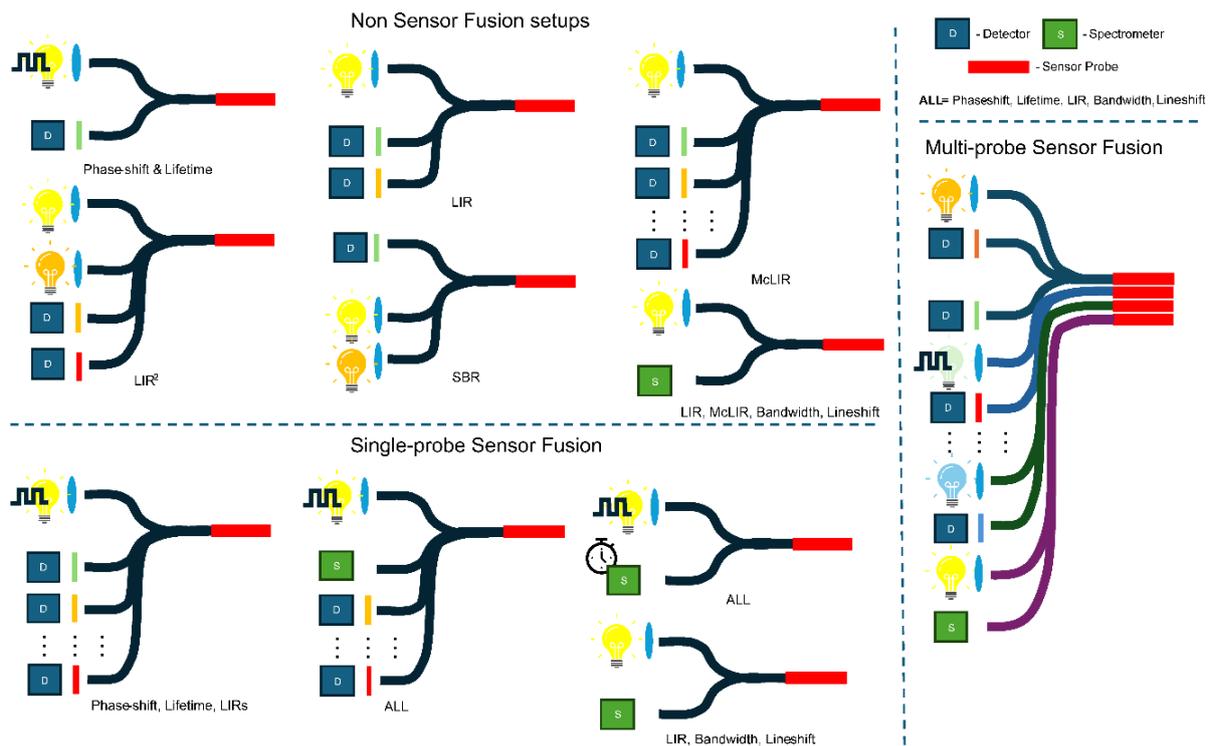

Figure 4. Experimental setups on an example of fiber-optic sensors and the readout method or combination for which they can be used.

The selection of sensor probes and parameters for SF depends on the specific goal of the SF method. SF can be used to extend the usable temperature range, as illustrated in Figure 5a, where different sensors dominate precision at various temperature ranges. In this scenario, SF functions more like an algorithm that switches between various sensors, without significantly increasing precision. If all sensors have maximum precision at the same temperature, SF will result in the increased precision within that temperature range, as shown in Figure 5b. With the proper selection of parameters and probes, SF can achieve both extended temperature range and increased precision. However, if one parameter or sensor is dominant across the entire temperature range, SF will not yield significant improvements.

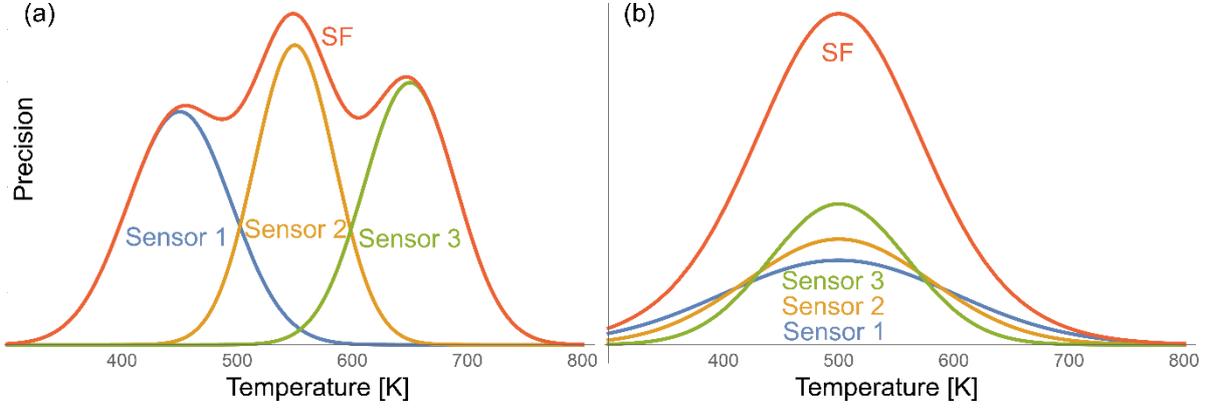

Figure 5. Combining 3 sensors by SF to increase (a) usable temperature range, (b) precision.

### 3.3. Sensor Fusion Luminescence Thermometry: Case studies

#### 3.3.1. LIR and line-shift of $Sm^{2+}$ doped $Al_2O_3$ coatings

Temperature dependent PL emission spectra under 450 nm excitation are presented in Figure 6a, showing two distinct spectral regions featured by the allowed 4f5d→4f and 4f→4f emissions. 4f5d and 4f levels of $Sm^{2+}$ are thermalized [43]. With increasing temperature the 4f5d→4f increase with temperature to the expense of 4f→4f transitions. 4f→4f transitions quickly quench with temperature, and they are barely visible after 600 K. At high temperatures those two emissions overlap, thus without deconvolution it is not possible to fit the LIR of 4f5d→4f and 4f→4f to the Boltzmann distribution equation. In the cases of the overlapping emissions, the Logistic curve excellently fits the experimental data:

$$y(x) = A + \frac{B}{1 + \exp\left(\frac{x - x_0}{d}\right)} \quad (12)$$

The ratio of emissions separated by 671 nm is fitted to the Logistic curve:

$$LIR = \frac{I_{525-671nm}}{I_{671-740nm}} = 3.17 - \frac{3.29}{1 + \exp\left(\frac{T - 501\ K}{69\ K}\right)} \quad (13)$$

The inverse of LIR is then:

$$T(LIR) = 501\ K + 69\ K \cdot \ln\left(-1 - \frac{3.29}{LIR - 3.17}\right) \quad (14)$$

The entire $Al_2O_3$:$Sm^{2+}$ emission barycenter energy was estimated from the Jacobi transformation of the spectra to the energy scale by [44]:

$$\nu[cm^{-1}] = \frac{10^7}{\lambda[nm]}, i_{cm^{-1}} = \frac{\lambda^2 i_{nm}}{10^7} \quad (15)$$

Then the barycenter energy was calculated by [45]:

$$\tilde{v}[\text{cm}^{-1}] = \frac{\int i_{\text{cm}^{-1}} v \, dv}{\int i_{\text{cm}^{-1}} dv} \quad (16)$$

The barycenter energy change of the entire emission spectra with temperature is also well fitted with the Logistic curve:

$$\tilde{v}[\text{cm}^{-1}] = 15571 \text{ cm}^{-1} - \frac{1621 \text{ cm}^{-1}}{1 + \exp\left(\frac{T - 435 \, K}{65 \, K}\right)} \quad (17)$$

Thus, the inverse of line-shift is:

$$T(\tilde{v}) = 435 \, K + 65 \, K \cdot \ln\left(-1 - \frac{1621 \text{ cm}^{-1}}{\tilde{v} - 15571 \text{ cm}^{-1}}\right) \quad (18)$$

And the fused temperature, uncertainty, and precision are:

$$\hat{T} = \sigma^2 \left(\frac{T_{LIR}}{\sigma_{LIR}^2} + \frac{T_v}{\sigma_v^2}\right), \sigma = \left(\frac{1}{\sigma_{LIR}^2} + \frac{1}{\sigma_v^2}\right)^{-1/2}, p = p_{LIR} + p_v \quad (19)$$

The precisions of LIR, line-shift, and SF are compared in Figure 6b.

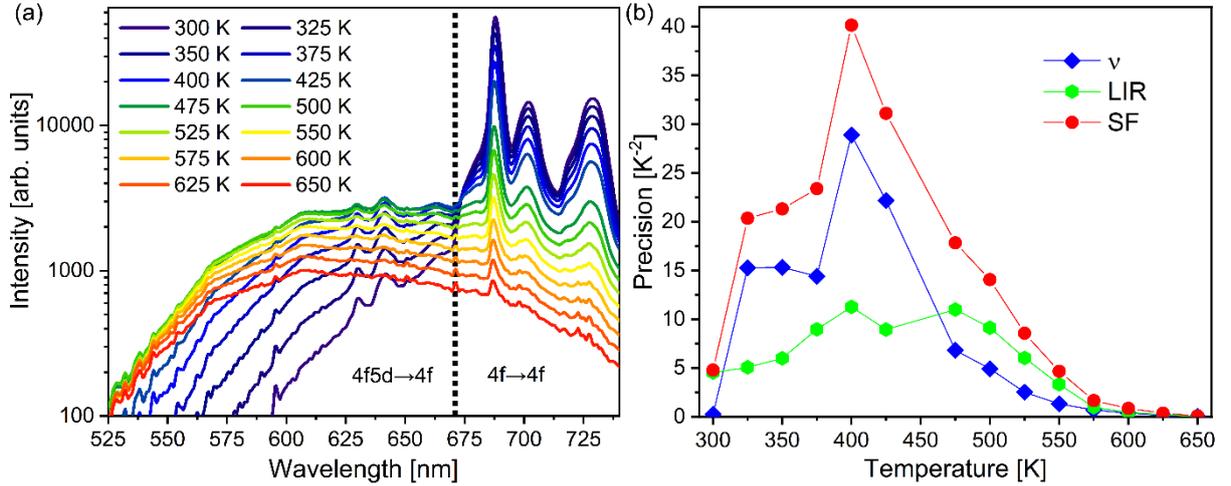

Figure 6. (a) PL emission of $Al_2O_3:Sm^{2+}$ coatings, excited with 450 nm laser. (b) Precision by line-shift, LIR, and SF methods.

### 3.3.2. Lifetime fusing with powder mixture of $Al_2O_3:Cr^{3+}$ + $Ba_3(PO_4)_2:Mn^{4+}$ + $Y_2O_3:Ho^{3+}$

The hypothesis for this powder combination is to achieve an extended usable temperature range. The ions were selected so that their main emissions do not spectrally overlap (see Figure 7a). $Cr^{3+}$ ion in strong crystal field exhibits sharp emission lines at about 700 nm. $Ho^{3+}$ has a sharp and intense green emission at about 550 nm. $Mn^{4+}$ position strongly depends on the host, and it is known to vary from 600 nm to 700 nm. Generally in fluoride hosts $Mn^{4+}$ emission is generally at higher energies than in oxide hosts [46,47], thus $Mn^{4+}$ doped fluorides are a good option for fusing with $Cr^{3+}$ emission.

It is known that $Mn^{4+}$ quenches with temperature more rapidly than $Cr^{3+}$, and transition metals quench much faster than lanthanide ions. To achieve the same intensity of $Cr^{3+}$, $Mn^{4+}$, and $Ho^{3+}$ emissions at room temperature, the powders were mixed in ratios of 2:3:9, respectively, before being pressed into a pellet. All lifetimes fit well with the Mott-Seitz relation. The inverse functions with fitted parameters are given by:

$$T_{\text{Cr}}(\tau) = \frac{2065 \text{ cm}^{-1}}{k \ln \frac{836.30\,\tau}{3.38\text{ ms} - \tau}}, T_{\text{Mn}}(\tau) = \frac{159 \text{ cm}^{-1}}{k \ln \frac{1.05\,\tau}{0.39\text{ ms} - \tau}}, T_{\text{Ho}}(\tau) = \frac{1605 \text{ cm}^{-1}}{k \ln \frac{40.18\,\tau}{0.049\text{ ms} - \tau}} \quad (20)$$

SF is then performed by:

$$\hat{T} = \sigma^2 \left( \frac{T_{\text{Cr}}}{\sigma_{\text{Cr}}^2} + \frac{T_{\text{Mn}}}{\sigma_{\text{Mn}}^2} + \frac{T_{\text{Ho}}}{\sigma_{\text{Ho}}^2} \right), \sigma = \left( \frac{1}{\sigma_{\text{Cr}}^2} + \frac{1}{\sigma_{\text{Mn}}^2} + \frac{1}{\sigma_{\text{Ho}}^2} \right)^{-1/2}, p = p_{\text{Cr}} + p_{\text{Mn}} + p_{\text{Ho}} \quad (21)$$

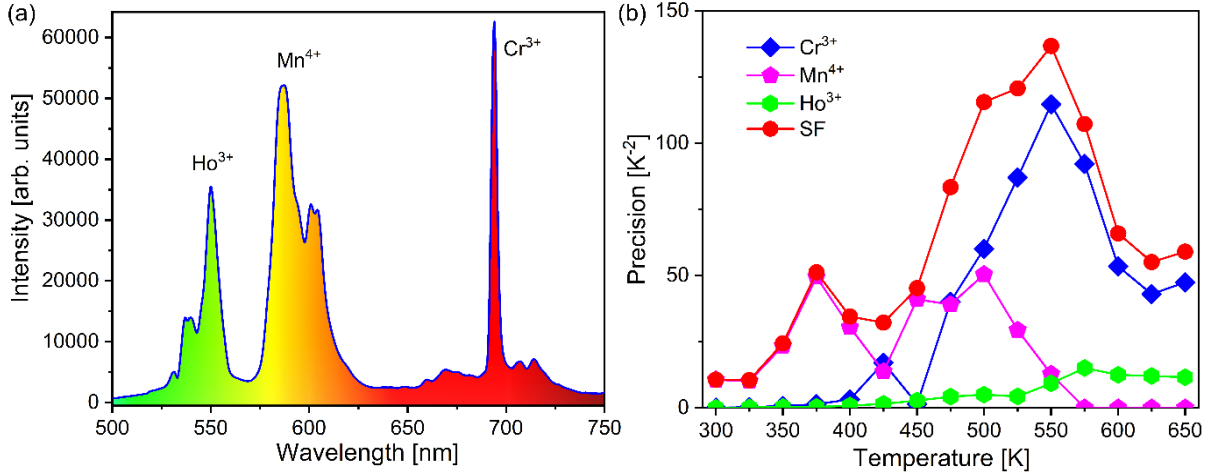

Figure 7. (a) Emission spectrum of $Al_2O_3$:$Cr^{3+}$, $Ba_3(PO_4)_2$:$Mn^{4+}$, and $Y_2O_3$:$Ho^{3+}$ powder mixture, and (b) precisions by their lifetimes and SF precision.

The results of the fused precision are presented in Figure 7b. As known for these activator ions, $Mn^{4+}$ is usable at the lowest temperatures, followed by $Cr^{3+}$ at intermediate, and then $Ho^{3+}$ at higher temperatures. $Mn^{4+}$ shows good precision from 350 K to 525 K. $Cr^{3+}$ can be used from 475 K up to 650 K, the end of the measured range. $Ho^{3+}$ becomes significant at temperatures above 550 K. At even higher temperatures than those measured here, $Ho^{3+}$ would become dominant as $Cr^{3+}$ emission would soon quench, whereas $Ho^{3+}$ would not.

The result of this combination of $Cr^{3+}$/$Mn^{4+}$/$Ho^{3+}$ mixed powder and single excitation is increased precision across the entire range, but more importantly, an extended usable temperature range compared to each individual ion.

### 3.3.3. Lifetime fusing of upconverting $YAlO_3$:$Yb^{3+}$/$Er^{3+}$ green and red emissions

After irradiation at 980 nm, $Yb^{3+}$ ions absorb the energy and transfer it to $Er^{3+}$ ions. Two energy transfers to $Er^{3+}$ are sufficient to create an upconversion emission, where the system absorbs lower-energy photons and generates higher-energy photons [48]. $Er^{3+}$ then emits green and red light from the $^4S_{3/2}$ and $^4F_{9/2}$ levels, respectively (see Figure 8a). The lifetimes of these emissions differ. Typically, a single emission is used for temperature sensing by lifetime measurement, as monitoring both complicates the fitting to a double-exponential relation. By using filters to separate the red and green emissions, we measured the lifetimes of each transition and fitted them to a single-exponential function. These measurements were performed rapidly to emulate sensing processes with fast changes (which increased measurement uncertainty). The following equations were used for SF:

$$\hat{T} = \sigma^2\left(\frac{T_S}{\sigma_S^2} + \frac{T_F}{\sigma_F^2}\right), \sigma = \left(\frac{1}{\sigma_S^2} + \frac{1}{\sigma_F^2}\right)^{-1/2}, p = p_S + p_F \quad (22)$$

The results are shown in Figure 8b. The lifetime of the $^4F_{9/2}$ level allows for more precise measurements compared to the $^4S_{3/2}$ level. However, the fused measurements showed slight but noticeable improvements in precision, demonstrating that monitoring multiple emissions from a single emission center with SF can enhance sensor probe performance.

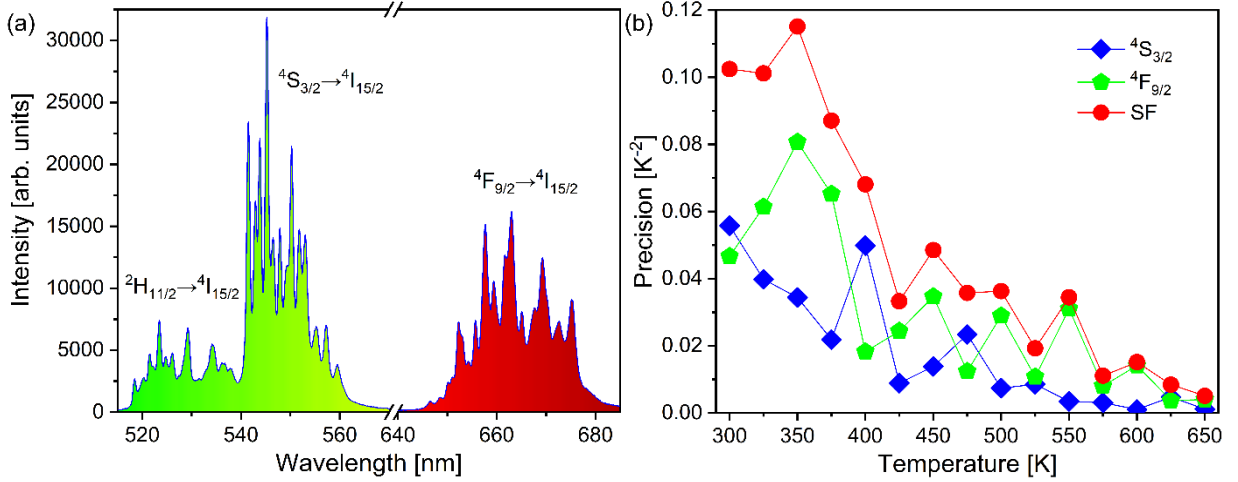

Figure 8. (a) Emission spectrum of YAM:$Yb^{3+}$/$Er^{3+}$ after 980 nm excitation. (b) Precision comparison of sensing lifetimes of red and green emissions of $Er^{3+}$ and their SF.

### 3.3.4. Two LIRs and line-shift of $Mn^{5+}$ doped $Ca_6BaP_4O_{17}$

The sensor fusion in luminescence thermometry was tested on an $Mn^{5+}$ doped sensor probe already tested for the MLR in Ref. [49]. Figure 9a shows the temperature dependent features of this probe, where 3 parameters are taken: (i) $LIR_1$ is the ratio of thermalized emissions from $^3T_2$ and $^1E$ to the ground level, (ii) $LIR_2$ is the ratio of anti-Stokes to Stokes sidebands of $^1E \rightarrow {}^3A_2$, and (iii) line-shift of $^1E \rightarrow {}^3A_2$ emission. The parameters are fitted to the following equations:

$$LIR_1 = B\exp\left(-\frac{\Delta E}{kT}\right) \Leftrightarrow T(LIR_1) = \frac{\Delta E}{k(\ln B - \ln LIR_1)} \quad (23)$$

$$LIR_2 = aT^2 + bT + c \Leftrightarrow T(LIR_2) = \frac{-b + \sqrt{b^2 - 4a(C - LIR_2)}}{2a} \quad (24)$$

$$\nu = d \cdot T + h \Leftrightarrow T(\nu) = \frac{\nu - h}{d} \quad (25)$$

The fused estimate of temperature, uncertainty, and precision are:

$$\hat{T} = \sigma^2\left(\frac{T_{LIR_1}}{\sigma_{LIR_1}^2} + \frac{T_{LIR_2}}{\sigma_{LIR_2}^2} + \frac{T_\nu}{\sigma_\nu^2}\right), \sigma = \left(\frac{1}{\sigma_{LIR_1}^2} + \frac{1}{\sigma_{LIR_2}^2} + \frac{1}{\sigma_\nu^2}\right)^{-1/2}, p = p_{LIR_1} + p_{LIR_2} + p_\nu \quad (26)$$

Due to the low intensity, $LIR_2$ had the highest uncertainty, thus its precision is much lower than for the other two parameters, and it could have been easily neglected in SF (see Figure 9b). Line-width demonstrated dominance in precision. $LIR_1$ had some contribution to the SF, improving the readout. However, this is the case where fusing multiple parameters lead to insignificant improvement in precision, demonstrating that for SF method it is desirable to have parameters with similar precision over the same temperature range in order to increase the precision. As line-shift has the largest precision over the entire measured temperature range, other parameters cannot be used in this probe for extending the temperature range.

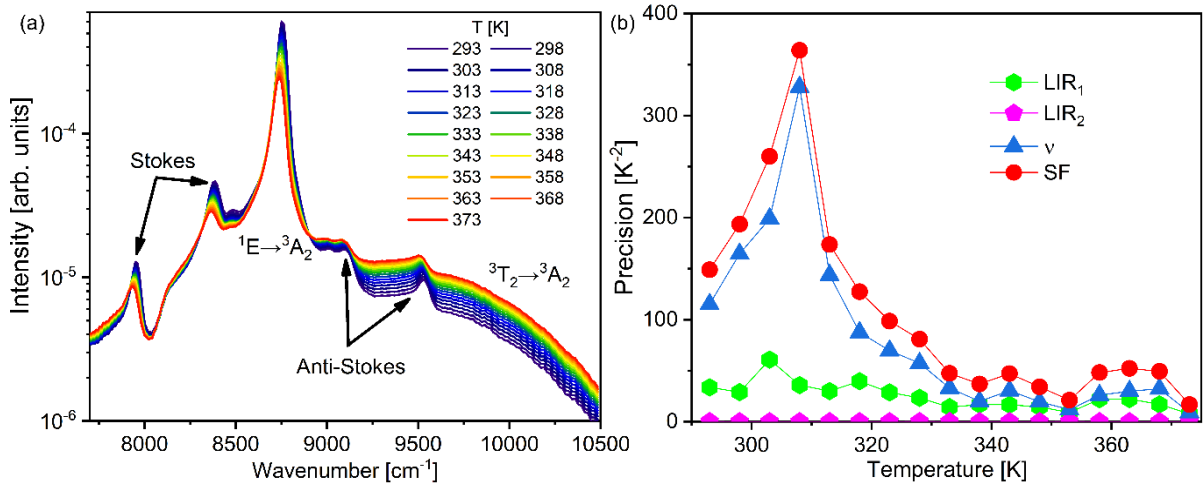

Figure 9. (a) Temperature-dependent spectra of $Ca_6BaP_4O_{17}$:$Mn^{5+}$ phosphor and (b) precision of each parameter, compared to the fused values.

## 4. Conclusion

Sensor Fusion (SF) luminescence thermometry, implemented through inverse variance weighting, demonstrates significant advantages by integrating multiple parameters that collectively enhance performance. This method is versatile, functioning effectively with a wide variety of parameters in both time-resolved and steady-state regimes, and is flexible enough to combine the two. SF can be established using either independent sensor probes or by treating individual parameters as distinct sensors.

A key consideration, however, is that if one parameter dominates in precision across the entire temperature range, the potential for SF to further improve precision is diminished compared to conventional thermometry. The most effective SF configurations are those where multiple parameters exhibit similar precision across a given range or when parameters complement each other by being more precise in different temperature ranges.

SF is both computationally and experimentally straightforward to implement. It requires only two essential equations: one for estimating fused uncertainty and another for determining the fused temperature. Experimentally, SF can be applied by either utilizing multiple probes at a single location or by optimizing experimental designs to exploit all measurable parameters from a single probe.

The demonstrated efficacy of SF in luminescence thermometry suggests that this approach may be extended to other luminescence-based sensing modalities, such as pressure or oxygen level measurements, opening pathways for broader applications.